\documentclass[a4paper,11pt]{article}
\usepackage{mathrsfs, amsmath, amsthm, amssymb}
\usepackage{citesort}
\usepackage{mcite}
\renewcommand{\d}{\textup{d}}

\begin{document}

\begin{titlepage}

\hfill QMUL-PH-11-15

\hfill DAMTP-2011-86

\vspace{0.75cm}
\begin{center}

{{\LARGE  \bf The local symmetries of M-theory and their formulation in generalised geometry}} \\

\vskip 1.5cm {David S. Berman$^\star$\footnote{D.S.Berman@qmul.ac.uk}, Hadi Godazgar$^\dag$\footnote{H.M.Godazgar@damtp.cam.ac.uk}, Mahdi Godazgar$^\dag$\footnote{M.M.Godazgar@damtp.cam.ac.uk} and Malcolm J. Perry$^{\dag}$\footnote{M.J.Perry@damtp.cam.ac.uk}}
\\
{\vskip 0.2cm
$^\star$Department of Physics,\\
Queen Mary University of London,\\
Mile End Road,
London,\\
E1 4NS, UK\\
}
{\vskip 0.2cm
$^\dag$DAMTP, Centre for Mathematical Sciences,\\
University of Cambridge,\\
Wilberforce Road, Cambridge, \\ CB3 0WA, UK\\}
\end{center}
\vskip 1 cm

\begin{center}
 \today
\end{center}

\vskip 1cm

\begin{abstract}
\baselineskip=18pt\
In the doubled field theory approach to string theory the T-duality group is promoted to a manifest symmetry at the expense of replacing ordinary Riemannian geometry with generalised geometry on a doubled space. The local symmetries are then given by a generalised Lie derivative and its associated algebra. This paper constructs an analogous structure for the extended geometry of M-theory. A crucial by-product of this construction is the derivation of the physical section condition for M-theory formulated in an extended space.
\end{abstract}

\end{titlepage}

\section{Introduction}
\noindent
The role of dualities in string and M-theory needs no introduction. Over the years there have been numerous attempts to make these duality symmetries \cite{SO(8), julia, TM, nicolaidewit, nicolia87, weste11} more manifest. In string theory this has led to the doubled geometry approach \cite{dft1, dft2, dft3, genmetstring} where the target space of the string is doubled and the metric on this space is given by the metric of generalised geometry of the sort developed by Hitchin and Gualtieri \cite{hitchin1, hitchin2, gualtieri}. This has various advantages: it makes the T-duality group a manifest symmetry and also naturally combines the metric and Kalb-Ramond field into a single {\it{geometric}} object known as the {\it{generalised metric}}. 

Of course one cannot just double the dimension of the space and remain with the same theory; string theory after all is notoriously tied to critical dimensions. One therefore needs a constraint which we refer to as the {\it{physical section condition}} that imposes that physics lives in a section of the doubled space. A solution of the section condition gives a submanifold of the original dimension. Different solutions of this section condition correspond to different duality frames. In this sense, duality is a spontaneously broken symmetry with solutions to the section condition breaking the duality invariance of the theory.

In this formulation, all the local symmetries of the theory may be naturally combined into a single generalised Lie derivative that would generate both ordinary diffeomorphisms and gauge transformations of the Kalb-Ramond field. The generalised Lie derivative generates a symmetry algebra, which after applying the section condition gives the Courant algebra. Since the Courant algebra is known to be the relevant algebra for generalised geometry \cite{courant}, it comes as no surprise to see its emergence from the combination of diffeomorphisms with gauge transformations \cite{hitchin1, hitchin2, gualtieri}.

One might suspect that this construction of string theory is somewhat artificial and alien to the usual formulation of string theory but in fact the structure of doubled geometry is present in a truncation of closed string field theory \cite{KZ}. In particular, apart from the inclusion of winding modes to provide the coordinates of the doubled space, closed string field theory gives the local symmetries, the subsequent Courant algebra and the section condition of doubled geometry first found by Siegel \cite{siegel1, siegel2}. From the string field theory perspective, the section condition is nothing more than an immediate consequence of level matching for the closed string.

The goal of this paper is to construct an analogous structure for M-theory. The extension of the space to give a linear realisation of the duality symmetry and the construction of a generalised metric on the extended space has already been given in a series of works \cite{hullm, PW, hillmann, davidmalcolm, BGP, BGPW1, BGPW2}. The construction in M-theory appears naturally and manifestly from the $E_{11}$ programme initiated by West and then by West and collaborators \cite{weste11, westl1, KW03, west04a, west04b, west10,westiia}. The relation between extended M-theory and doubled geometry was described in detail in \cite{relSM}. Yet in these works there were some crucial open questions. Namely, what are the local symmetries of the extended theory, i.e.\ what is the {\it{generalised}} Lie derivative for this {\it{generalised geometry}}? What is the algebra of this local symmetry? Perhaps most importantly, what is the {\it{physical section condition}}?

This paper answers these questions. The method is as follows. The local symmetry is constructed so as to produce ordinary diffeomorphisms and gauge transformations once a duality frame is picked. This produces the {\it{generalised Lie derivative}} on the space. The Courant bracket is defined to be the antisymmetrisation of the generalised Lie derivative. The algebra is calculated and shown to be a Courant algebra provided a particular condition quadratic in derivatives is obeyed. This condition will then be taken to be the {\it{physical section condition}}. Consistency is then checked by comparing with the string result through dimensional reduction.

Recent, related work include the extension of double field theory to Heterotic string theory \cite{dhet} and type II strings \cite{dtypeii, dtypeiim, CSW} and other interesting studies given in \cite{Albertsson:2008gq, Jeon:2010rw, Hohm:2010xe, Jeon:2011kp, Jeon:2011cn, Andriot:2011uh, Albertsson:2011ux, Kan:2011vg, Aldazabal:2011nj, Jeon:2011vx, boundary}. 

\section{Generalised Lie derivative and Courant Bracket}
\label{geo}

The SL(5) duality group is found in the reduction of eleven-dimensional supergravity to seven dimensions. In \cite{davidmalcolm}, this duality group is made manifest in the directions of the 4-torus without assuming the existence of isometries. Since the duality group is made to act along four directions, the only supergravity fields that are non-vanishing are the metric and the 3-form potential $C.$ The corresponding generalised geometry involves the extension of vectors, making up the tangent space, by 2-forms. The two-forms are the windings of the membrane, which source $C$. Furthermore, it was shown in \cite{davidmalcolm} that the diffeomorphism and gauge symmetry of the 3-form potential are a result of reparametrisations of the ordinary space coordinates and winding coordinates, respectively. These form a Courant bracket algebra, which is exactly the same algebra found in generalised geometry \cite{courant, hitchin1, gualtieri}. 

The analogue of a Lie derivative in generalised geometry is a generalised Lie derivative \cite{GMPW}, which encodes reparametrisations of the generalised coordinates as well as ordinary coordinate transformations. For generalised vector fields $V=(v, \mu)$ and $X=(x, \lambda),$ where $v$ and $x$ are vector fields and $\mu$ and $\lambda$ are 2-forms, the generalised Lie derivative is defined to be 
\begin{equation}
 \hat{\mathcal{L}}_{X} V = \mathcal{L}_{x} v + \mathcal{L}_{x} \mu - i_{v} \d \lambda.
\label{defgenLie}
\end{equation}
The first term is the Lie derivative of the vector field $v$ along $x.$ This reproduces the transformation of a vector field under coordinate transformations. The second and third terms in the generalised Lie derivative give the transformation of a two-form field, $\mu,$ under coordinate transformations generated by $x$ and gauge transformations generated by $\lambda.$ 

The antisymmetrisation of the generalised Lie derivative gives rise to a Courant bracket, as opposed to a Lie bracket,
\begin{align}
[X,Y]_{C} &= \frac{1}{2} \left(\hat{\mathcal{L}}_{X} Y - \hat{\mathcal{L}}_{Y} X\right) \label{defC}\\
&= [x,y] + \mathcal{L}_{x} \eta - \mathcal{L}_{y} \lambda +\frac{1}{2} \d \left( i_{y} \lambda - i_{x} \eta \right),
\label{Courant}
\end{align}
where $X=(x,\lambda), \ \ Y=(y, \eta)$ and $[x,y]$ denotes the Lie bracket of vector fields $x$ and $y.$ 
One may view the Courant bracket as describing the algebra of combined diffeomorphisms and gauge transformations.
Thus the algebra of diffeomorphisms is exactly as one would expect as can be seen from the first term 
on the right-hand side of \eqref{Courant} and is given by the Lie bracket. The second and third terms 
on the right-hand side of \eqref{Courant} are what one would expect from  a gauge transformation
followed  by a diffeomorphism. The last term is perhaps a little surprising in that it is exact. A gauge
transformation that is exact will have no effect on the three-form potential $C$.

Given this, one might wonder if the Jacobi transformations hold. The Jacobiator of the Courant bracket 
is defined by
\begin{equation}
J(X,Y,Z) = [[X,Y]_C,Z]_C + [[Y,Z]_C,X]_C + [[Z,X]_C,Y]_C
\end{equation}
and measures by how much the Jacobi identities fail. Using $X=(x,\lambda), Y=(y,\mu)$ and $Z=(z,\kappa)$
we find 
\begin{equation}
J(X,Y,Z) = \frac{1}{4}d[(\iota_xL_y-\iota_yL_x)\kappa + (\iota_yL_z-\iota_zL_y)\lambda +(\iota_zL_x-\iota_xL_z)\mu].
\end{equation}
Since $J$ is exact, the Jacobi identity holds when restrcited to being evaluated on the field $C$ which is the only field that
changes under a gauge transformation. It trivially holds on all the other fields that do not transform under 
this gauge trnasformation. 

The generalised vector fields can be twisted by a 3-form $C$ in the following way:
$$ \rho_{C}(X,\lambda)= (x,\lambda + \frac{1}{\sqrt{2}} i_{X} C).$$
The Courant bracket in terms of the twisted vector fields now reproduces the algebra of diffeomorphisms and gauge symmetries \cite{davidmalcolm}.
The generalised Lie derivative and Courant bracket above, equations \eqref{defgenLie} and \eqref{Courant} respectively, treat each ``component'' of the generalised vector field separately. The distinction made between the coordinates and windings is unnatural from the perspective of making duality a manifest symmetry. We would like a more democratic, or covariant, formulation of these objects. This  allows us to find the section condition for the generalised geometry of the SL(5) duality group. 

In components, the generalised vector field $X$ is $(x^{i}, \mu_{ij}),$ where $i,j=1,\dots,4.$ The 2-form can be Hodge dualised with the alternating symbol $\eta$ ($\eta^{1234}=1$) so that the indices on $X$ become SL(5) indices, viz.\
\begin{equation} \label{genfield}
 X^{a b}=
\begin{cases}
X^{i5} = x^{i} \\
X^{5i} = -x^{i} \\
X^{ij} = \frac{1}{2} \eta^{ijkl} \mu_{kl}
\end{cases},
\end{equation}
where $a=(i,5).$ The generalised Lie derivative, defined in equation \eqref{defgenLie}, can be written as two pieces with different tensor structures
\begin{gather}
 \left(\hat{\mathcal{L}}_{X} V \right)^{i} = x^{k} \partial_{k} v^{i} - v^{k} \partial_{k} x^{i}, \label{genLiei} \\
  \left(\hat{\mathcal{L}}_{X} V \right)_{ij} = x^{k} \partial_{k} \mu_{ij} + \mu_{ik} \partial_{j} x^{k} + \mu_{kj} \partial_{i} x^{k} - 3 v^{k} \partial_{[k} \lambda_{ij]}.
\end{gather}
Now, the Hodge dual of the second equation above is 
\begin{align}
  \left(\hat{\mathcal{L}}_{X} V \right)^{ij} &=  \frac{1}{2} \eta^{ijkl} \left( x^{m} \partial_{m} \mu_{kl} + \mu_{km} \partial_{l} x^{m} + \mu_{ml} \partial_{k} x^{m} - 3 v^{m} \partial_{[m} \lambda_{kl]} \right) \notag \\
&= x^{m} \partial_{m} V^{ij} + 2 V^{k[i} \partial_{k}x^{j]} + V^{ij} \partial_{k} x^{k} + 2 v^{[i} \partial_{k} X^{j]k},
\label{genLieij}
\end{align}
where $V^{ij}= \frac{1}{2} \eta^{ijkl} \mu_{kl} $ and similarly $X^{ij}= \frac{1}{2} \eta^{ijkl} \lambda_{kl}.$ We would like to write the generalised Lie derivative in terms of indices $a, b$ that run from 1 to 5, $ (\hat{\mathcal{L}}_{X} V)^{ab},$ so that if $b$ is 5 we get the expression on the right-hand side of 
equation \eqref{genLiei} and if $a,b$ take values from 1 to 4 then we get the expression on the right-hand side of equation \eqref{genLieij}. The expression that reduces to equations \eqref{genLiei} and \eqref{genLieij} is 
\begin{equation}
 (\hat{\mathcal{L}}_{X} V)^{ab} = \frac{1}{2} X^{cd} \partial_{cd} V^{ab} + \frac{1}{2} V^{ab} \partial_{cd} X^{cd} + V^{ac} \partial_{cd} X^{db} - V^{cb} \partial_{cd} X^{ad},
\label{genLiecov}
\end{equation}
assuming that $\partial_{ij} =0.$ Although to write the generalised Lie derivative in terms of the generalised fields we assumed that $\partial_{ij} =0,$ in what follows we drop this condition. The requirement that $\partial_{ij} =0$ is one particular choice to make the restriction from the 10-dimensional extended space to the four-dimensional physical space. This is one solution of the section condition, but there are other choices that can be made. 

The generalised Lie bracket that reduces to the Courant bracket, equation \eqref{Courant}, when particular components are considered is, as in the case of the Lie or Courant bracket, the antisymmetrisation of the corresponding Lie derivative, \eqref{defC}. Therefore, antisymmetrising the generalised Lie derivative, defined in equation \eqref{genLiecov}, gives the generalised Lie bracket
\begin{gather}
 [X,Y]^{ab}_{G} = \frac{1}{4} X^{cd} \partial_{cd} Y^{ab} - \frac{1}{4} Y^{cd} \partial_{cd} X^{ab}+ X^{[a|c} \partial_{cd} Y^{|b]d}  \notag \\ 
\qquad \qquad \qquad \qquad \qquad - Y^{[a|c} \partial_{cd} X^{|b]d} - \frac{1}{4} X^{ab} \partial_{cd} Y^{cd} + \frac{1}{4} Y^{ab} \partial_{cd} X^{cd}.
\label{genbra}
\end{gather}
For particular choices of indices $a,b,$ the above reduces to the Courant bracket, equation \eqref{Courant}. For example, letting $a=i$ and $b=5$ in the above we get $[x,y]^i,$ which agrees with the Courant bracket. A similar check can easily be done for the choice $ab=ij.$ If we think of $a,b$ as ordinary coordinate indices, then we should only get ordinary differential geometry. In the case of Riemannian geometry one should only find the first two terms, which are the Lie bracket of $X$ and $Y.$

The Jacobi identity for the generalised Lie bracket is not satisfied. However, the Jacobi identity holds on fields up to terms that vanish by the section condition (see appendix \ref{Jac}). Physically, the Jacobiator can be shown to be a pure gauge transformation on fields.

The algebra of generalised diffeomorphisms, which includes diffeomorphisms and gauge symmetries, must be closed. However, we find that
\begin{align}
 \left( [\hat{\mathcal{L}}_{X},\hat{\mathcal{L}}_{Y}] V \right)^{ab} &= \left( \hat{\mathcal{L}}_{[X,Y]_{G}} V \right) ^{ab} + \frac{3}{4} \biggl[ \biggl( X^{ef} \partial_{[ef} Y^{cd} \cdot \partial_{cd]} V^{ab} \notag \\ 
& \qquad +  V^{ab} X^{ef} \partial_{[ef}  \partial_{cd]} Y^{cd} + V^{ab} \partial_{[ef} X^{ef} \cdot \partial_{cd]} Y^{cd} \notag \\ 
& \qquad \quad - 4 V^{c[a} \partial_{[ef}  Y^{b]d} \cdot \partial_{cd]} X^{ef}  + 2 X^{ef} V^{c[a} \partial_{[ef} \partial_{cd]} Y^{b]d} \notag \\
& \qquad \qquad  + 2 V^{c[a} X^{b]d} \partial_{[ef} \partial_{cd]} Y^{ef} \biggr) - (X \leftrightarrow Y) \biggr].
\label{antiLie}
\end{align}
Since closure of the algebra requires that 
$$[\hat{\mathcal{L}}_{X},\hat{\mathcal{L}}_{Y}] V = \hat{\mathcal{L}}_{[X,Y]_{G}} V, $$
the rest of the terms on the right-hand side of equation \eqref{antiLie} must vanish which leads us to the section condition for the SL(5) generalised geometry. All of the extra terms vanish if 
\begin{equation}
 \partial_{[ab} \partial_{cd]} X =0 \qquad \text{and} \qquad  \partial_{[ab} X \cdot \partial_{cd]} Y  =0,
\label{seccon}
\end{equation}
where $X$ and $Y$ are arbitrary fields. This is the section condition that has to be satisfied by the generalised fields in the SL(5) generalised geometry.  Letting the fields depend only on the $x^{i}$ coordinates and imposing $$ \frac{\partial}{\partial y_{ij}} =0 $$ on all fields is one solution to the section condition, which was the one considered in \cite{davidmalcolm}, but there are also other possibilities. Solutions to the section condition will be considered in section \ref{sectioncon}.

The generalised Lie derivative has been defined with respect to the antisymmetric representation of SL(5). We can in principle write the generalised Lie derivative of a field in any representation of SL(5), including the fundamental representation. This is analogous to the spinorial Lie derivative in differential geometry.  For the generalised field in the fundamental representation the generalised Lie derivative takes the form
\begin{equation}
 (\hat{\mathcal{L}}_{X} V)^{a} = \frac{1}{2} X^{cd} \partial_{cd} V^{a} + \frac{1}{4} V^{a} \partial_{cd} X^{cd}  - V^{c} \partial_{cd} X^{ad}.
\label{genLiecovf}
\end{equation}
Evaluating the generalised Lie derivative of $U^{[a} V^{b]}$ using the above equation one can easily see that this definition is consistent with the generalised Lie derivative of a field in the antisymmetric representation.

The generalised Lie derivative of other SL(5) objects can be computed by a simple application of the Leibniz rule and by assuming that the generalised Lie derivative on a scalar is equal to its partial derivative, i.e.
\begin{equation}
  \hat{\mathcal{L}}_{X} S=\frac{1}{2} X^{cd} \partial_{cd} S.
\end{equation}
For example, the generalised Lie derivative on an SL(5) covector $V_{a}$ is
\begin{equation} \label{genLiesdown}
(\hat{\mathcal{L}}_{X} V)_{a} = \frac{1}{2} X^{cd} \partial_{cd} V_{a} - \frac{1}{4} V_{a} \partial_{cd} X^{cd} + V_{c} \partial_{ad} X^{cd}.
\end{equation}
Or, the generalised Lie derivative on an antisymmetric object $V_{ab}$ is
\begin{equation} \label{genLievdown}
(\hat{\mathcal{L}}_{X} V)_{ab}=\frac{1}{2} X^{cd} \partial_{cd} V_{ab} - \frac{1}{2} V_{ab} \partial_{cd} X^{cd} + V_{a d} \partial_{c b} X^{cd} - V_{b d} \partial_{c a} X^{cd}.
\end{equation}

The generalised Lie derivative is supposed to encode all gauge freedoms associated with the theory. In particular, one would expect the generalised Lie derivative on the generalised metric to induce diffeomorphism and gauge transformations of the metric and 3-form potential.

The generalised metric for the SL(5) generalised geometry given in \cite{davidmalcolm} is
$$\begin{pmatrix} g_{ij} + \frac{1}{2} {C_{i}}^{kl} C_{jkl} & \frac{1}{\sqrt{2}} {C_{i}}^{\;kl} \\
 \frac{1}{\sqrt{2}} {C^{mn}}_{j} & g^{mn,kl} \end{pmatrix},$$
where $g_{ij}$ is the metric, $C_{ijk}$ is the 3-form potential and $g^{mn,kl}=\frac{1}{2}(g^{mk}g^{nl}-g^{ml}g^{nk})$.
However, in order to make connection with the SL(5) group, we dualised the winding coordinates and combined them with the ordinary coordinates to construct a object, $X^{ab},$ with SL(5) indices, equation \eqref{genfield}. The metric that acts on $X^{ab},$ or equivalently the $\mathbf{10}$ of SL(5), is the generalised metric\footnote{In terms of the M2-brane picture from which the original generalised metric was derived in \cite{davidmalcolm}, this generalised metric can be derived by considering the Hodge dual of the Lagrange multiplier used in the membrane action.} 
\begin{equation} \label{bigmet}
M_{ab, \, cd}= \begin{pmatrix} M_{i5, \, j5} & M_{i5, \, mn} \\
M_{kl, \, j5} & M_{kl, \, mn} \end{pmatrix}
= \begin{pmatrix} g_{ij} + \frac{1}{2} {C_{i}}^{pq} C_{jpq} & - \frac{1}{2 \sqrt{2}} {C_{i}}^{pq} \eta_{pqmn} \\
- \frac{1}{2\sqrt{2}} C_{j}^{\;pq} \eta_{pqkl} & g^{-1} g_{kl,mn} \end{pmatrix},
\end{equation}
where $g=\det(g_{ij})$ and $g_{kl,mn}=\frac{1}{2}(g_{mk}g_{nl}-g_{ml}g_{nk})$ so that $g^{kl,mn}g_{mn,pq}=\frac{1}{2}(\delta^{k}_{p} \delta^{l}_{q}- \delta^{k}_{q} \delta^{l}_{p}).$

The generalised metric above is determined by $g_{ij}$ and $C_{ijk}.$ Hence, it has $10+4=14$ number of independent components. A general symmetric $10 \times10$ matrix has 55 independent components. However, the metric $M$ is constrained to parametrise the coset SL(5)/SO(5), which is a $24-10=14$ dimensional space. 

Using equation \eqref{genLievdown}, we find that under a generalised Lie derivative, the variation of the generalised metric is 
\begin{align}
 (\hat{\mathcal{L}}_{X} M)_{ab, \, cd} =& \frac{1}{2} X^{ef} \partial_{ef} M_{ab, \, cd} -  M_{ab, \, cd} \partial_{ef} X^{ef} \notag \\ & + M_{a f, \, cd} \partial_{e b} X^{ef} - M_{b f, \, cd} \partial_{e a} X^{ef} \notag \\ & \hspace{5mm} + M_{a b, \, cf} \partial_{e d} X^{ef} - M_{ab, \, df} \partial_{e c} X^{ef}.
\end{align}
In particular, taking the $ab=i5$, $cd=j5$ components gives
\begin{align}
 (\hat{\mathcal{L}}_{X} M)_{i5, \, j5} =& x^{k} \partial_{k} (g_{ij} + \textstyle{\frac{1}{2}} {C_{i}}^{pq} C_{jpq}) 
+(g_{kj} + \textstyle{\frac{1}{2}} {C_{k}}^{pq} C_{jpq}) \partial_{i} x^k \notag \\ & \hspace{45mm}+(g_{ik}  + \frac{1}{2} {C_{i}}^{pq} C_{kpq}) \partial_{j} x^k
\notag \\
&- \frac{1}{2 \sqrt{2}} \partial_{k} X^{kl} ({C_{i}}^{pq} \eta_{pq jl} +{C_{j}}^{pq} \eta_{pq il}) + \ldots,
\end{align}
where we have used equation \eqref{genfield}, notably that $X^{i5}=x^i,$ and the form of the generalised metric given in \eqref{bigmet}. The ellipses denote terms involving derivatives with respect to winding coordinates, i.e. $\partial_{ij}$, which we are not interested in.

Letting $X^{ij}= - \frac{1}{\sqrt{2}} \eta^{ijkl} \Lambda_{kl}$, the expression above reduces to
\begin{equation}
  (\hat{\mathcal{L}}_{X} M)_{i5, \, j5} = \mathcal{L}_x (g_{ij} + \textstyle{\frac{1}{2}} {C_{i}}^{pq} C_{jpq}) + \textstyle{\frac{1}{2}} ({C_{i}}^{pq} \partial_{[j}\Lambda_{pq]} +{C_{j}}^{pq} \partial_{[i}\Lambda_{pq]}) + \ldots,
\end{equation}
allowing us to conclude that the generalised Lie derivative does indeed induce diffeomorphism and gauge transformations in a manner in which we would expect it to.  Similar computations on the other components of the generalised metric result in the expected transformations.

The form of the generalised metric in equation \eqref{bigmet} allows us to write it as a pair of objects acting on the \textbf{5} of SL(5). That is,
\begin{equation} \label{biglit}
 M_{ab, \, cd} = m_{ac} m_{bd} - m_{ad} m_{bc},
\end{equation}
where the symmetric metric $m$ is
\begin{equation} \label{litmet}
m_{ab}= \begin{pmatrix} g^{-1/2} g_{ij} & V_{i} \\
V_{j} & g^{1/2} (1+g_{ij} V^{i} V^{j}) \end{pmatrix}.
\end{equation}
The SL(4) vector
\begin{equation}
 V^{i}=\frac{1}{6} \epsilon^{ijkl} C_{jkl},
\end{equation}
where $\epsilon^{ijkl}= g^{-1/2} \eta^{ijkl}$ and is the alternating tensor. The fact that $M$ can be written as in equation \eqref{biglit} is related to the generalised metric parametrising the coset space SL(5)/SO(5) and can be thought of as the constraint that reduces the number of degrees of freedom of $M$ from 55 to 14. 

A natural question to consider is whether the structure in equation \eqref{biglit} is preserved under generalised diffeomorphisms.  In other words, does the metric $m$ transform as one would expect its components to transform.  The generalised Lie derivative of $m$ is
\begin{equation} \label{litmettran}
 (\hat{\mathcal{L}}_{X} m)_{ab} = \frac{1}{2} X^{cd} \partial_{cd} m_{ab} - \frac{1}{2} m_{ab} \partial_{cd} X^{cd} + m_{cb} \partial_{ad} X^{cd} + m_{ac} \partial_{bd} X^{cd}.
\end{equation}
Taking the $a=i$, $b=j$ component of the equation above gives
\begin{align}
  (\hat{\mathcal{L}}_{X} m)_{ij} = & x^{k} \partial_{k} (g^{-1/2} g_{ij}) + g^{-1/2} g_{kj} \partial_{i} x^{k} 
+ g^{-1/2} g_{ik} \partial_{j} x^{k} - g^{-1/2} g_{ij} \partial_{k} x^{k} + \ldots \notag \\
= & \mathcal{L}_x( g^{-1/2} g_{ij}) + \ldots.
\end{align}
Hence, the generalised Lie derivative on $m$ reproduces diffeomorphisms.
Now, taking $a=i$, $b=5$,
\begin{align}
 (\hat{\mathcal{L}}_{X} m)_{i5}= & x^{k} \partial_{k} (V_{i}) + V_{k} \partial_{i} x^{k} - g^{-1/2} g_{ik} \partial_{l} X^{kl} + \ldots \notag \\
= & (\mathcal{L}_x V)_{i} + \delta_{\Lambda} V_{i} + \ldots,
\end{align}
where
\begin{equation}
 \delta_{\Lambda} V_{i} = \frac{1}{6} g_{ik} \epsilon^{klmn} \delta_{\Lambda} C_{lmn} = \frac{1}{2} g_{ik} \epsilon^{klmn} \partial_{l} \Lambda_{mn} = - g^{-1/2} g_{ik} \partial_{l} X^{kl},
\end{equation}
for $X^{kl} = - \textstyle{\frac{1}{2}} \eta^{klmn} \Lambda_{mn}.$ A similar result is obtained if we choose the $5\,5$ component of equation \eqref{litmettran}. We conclude that equation \eqref{biglit} is preserved under generalised diffeomorphisms. In other words, generalised diffeomorphisms preserve the coset structure in which $M$ lies.

\section{Relation to O$(d,d)$ generalised geometry}
\label{rel33}

In this section, the generalised Lie derivative and bracket of the SL(5) generalised geometry, \eqref{genLiecov} and \eqref{genbra}, are related to the corresponding objects of the O$(d,d)$ generalised geometry \cite{siegel1, siegel2, genmetstring, CSW}. The dimensional reduction of the SL(5) duality manifest description of M-theory should be related to the O(3,3) structure in string theory. Therefore, by dimensional reduction of the objects describing the SL(5) generalised geometry we should recover the O(3,3) generalised geometry. 

First, let us consider the generalised Lie derivative. We reduce along the $a=4$ direction, so we let any derivative with 4 as an index vanish, and consider the fields along the first three directions which we label by  $\alpha, \beta , \dots = 1,2,3.$ Let $a= \alpha$ and $b=5$ in equation \eqref{genLiecov},
\begin{align*}
(\hat{\mathcal{L}}_{X} V)^{\alpha 5} = \frac{1}{2} X^{cd} \partial_{cd} V^{\alpha 5} + \frac{1}{2} V^{\alpha 5} \partial_{cd} X^{cd} - V^{\alpha c} \partial_{cd} X^{5 d} - V^{c 5} \partial_{c d} X^{\alpha d}.
\end{align*}
Note that because of the reduction ansatz the indices on the derivatives $\partial_{cd}$ can only take values along the first three directions or the fifth direction, viz. $c,d = \gamma$ or $5,$ but because of the antisymmetry both $c$ and $d$ cannot be 5. Hence, denoting 
$$V^{\alpha 5} = V^{\alpha}, \quad X^{\alpha 5} = X^{\alpha} \quad \textup{ and } \quad \partial_{\alpha 5} = \partial_{\alpha},$$
write 
\begin{align*}
(\hat{\mathcal{L}}_{X} V)^{\alpha} = & X^{\gamma} \partial_{\gamma} V^{\alpha} + \frac{1}{2} X^{\gamma \delta} \partial_{\gamma \delta} V^{\alpha} + V^{\alpha} \partial_{\gamma} X^{\gamma}+ \frac{1}{2} V^{\alpha} \partial_{\gamma \delta} X^{\gamma \delta}\\
& \qquad - V^{\alpha 5} \partial_{5 \gamma} X^{5 \gamma} - V^{\alpha \gamma} \partial_{\gamma \delta} X^{5 \delta} - V^{\gamma 5} \partial_{\gamma 5} X^{\alpha 5}- V^{\gamma 5} \partial_{\gamma \delta} X^{\alpha \delta} \\
= & X^{\gamma} \partial_{\gamma} V^{\alpha} + \frac{1}{2} X^{\gamma \delta} \partial_{\gamma \delta} V^{\alpha} + \frac{1}{2} V^{\alpha} \partial_{\gamma \delta} X^{\gamma \delta}\\
& \qquad + V^{\alpha \gamma} \partial_{\gamma \delta} X^{\delta} - V^{\gamma} \partial_{\gamma} X^{\alpha}- V^{\gamma} \partial_{\gamma \delta} X^{\alpha \delta}.
\end{align*}
Identifying 
$$\tilde{V}_{\alpha} = \frac{1}{2} \eta_{\alpha \beta \gamma} V^{\beta \gamma}, \quad \tilde{X}_{\alpha} = \frac{1}{2} \eta_{\alpha \beta \gamma} X^{\beta \gamma} \quad \textup{ and } \quad \tilde{\partial}^{\alpha} = \frac{1}{2} \eta^{\alpha \beta \gamma} \partial_{\beta \gamma},$$
where $\eta_{\alpha \beta \gamma}$ is the alternating symbol, the expression for the O(3,3) generalised Lie derivative acting on a vector $V^{\alpha}$ is
\begin{align}
(\hat{\mathcal{L}}_{X} V)^{\alpha} = & X^{\gamma} \partial_{\gamma} V^{\alpha} + \tilde{X}_{\gamma} \tilde{\partial}^{\gamma} V^{\alpha} - \tilde{V}_{\gamma} \tilde{\partial}^{\gamma} X^{\alpha}+ \tilde{V}_{\gamma} \tilde{\partial}^{\alpha} X^{\gamma} - V^{\gamma} \partial_{\gamma} X^{\alpha}+ V^{\gamma}  \tilde{\partial}^{\gamma} \tilde{X}_{\gamma} \notag \\
=& X^{\Pi} \partial_{\Pi} V^{\alpha}- V^{\Pi} \partial_{\Pi} X^{\alpha} + V^{\Pi} \tilde{\partial}^{\alpha} X_{\Pi},
\label{o33u}
\end{align}
where 
$$ X^{\Pi}= (X^{\alpha}, \tilde{X}_{\alpha}), \qquad X_{\Pi}= (\tilde{X}_{\alpha}, X^{\alpha})$$ and
$$\partial_{\Pi}= (\partial_{\alpha}, \tilde{\partial}^{\alpha}), \qquad \partial^{\Pi}= (\tilde{\partial}^{\alpha}, \partial_{\alpha}).$$
Similarly, we find the O(3,3) generalised Lie derivative acting on a field with lowered indices by considering the Hodge dual of equation \eqref{genLiecov} with  $a= \beta$ and $b=\gamma,$
\begin{align}
(\hat{\mathcal{L}}_{X} \tilde{V})_{\alpha} = & \frac{1}{2} \eta_{\alpha \beta \gamma} (\hat{\mathcal{L}}_{X} V)^{\beta \gamma} \notag \\
=& X^{\Pi} \partial_{\Pi} \tilde{V}_{\alpha}- V^{\Pi} \partial_{\Pi} \tilde{X}^{\alpha} + V^{\Pi} \partial_{\alpha} X_{\Pi}.
\label{o33d}
\end{align}
Therefore, putting together equations \eqref{o33u} and \eqref{o33d} we find the O(3,3) covariant generalised Lie derivative given in equation (3.22) of \cite{genmetstring},
$$(\hat{\mathcal{L}}_{X} V)^{\Sigma} =  X^{\Pi} \partial_{\Pi} V^{\Sigma}- V^{\Pi} \partial_{\Pi} X^{\Sigma} + V^{\Pi} \partial^{\Sigma} X_{\Pi}.$$

The dimensional reduction of the generalised Lie derivative on a generalised vector field in the antisymmetric representation of SL(5) corresponds to the generalised Lie derivative on a generalised vector field in the fundamental of O(3,3). Therefore, one would expect that dimensionally reducing the expression for the generalised Lie derivative on a generalised field in the fundamental representation of SL(5) is related to the generalised Lie derivative of a generalised field in some other representation of O(3,3). This is indeed the case as we will now show. A field in the antisymmetric representation, $W^{ab},$ can be constructed from fields in the fundamental representation, $U^{a}$ and $V^{a},$ by simply taking the antisymmetrisation of the two fields,
$$ W^{ab}= U^{[a} V^{b]}.$$ We showed that under dimensional reduction $W^{ab}$ becomes a field in the vector of O(3,3), which we denote $W^{\Pi}.$ When $a= \alpha$ and $b=5,$ we have the following picture
\medskip
$$
\begin{array}{ccccccccccccccc}
W^{ab} & = & U^{[a} V^{b]} \\
\downarrow && \downarrow \\
W^{\alpha} && U^{[\alpha}V^{5]}
\end{array}.
$$
\medskip

The above diagram shows that the O(3,3) vector field $W^{\alpha}$ must also be decomposable in terms of fields $U$ and $V$ that are in a lower dimensional representation of O(3,3). The only non-trivial representation of O(3,3) that is possible is the Majorana-Weyl representation of O(3,3)\footnote{More technically, it is the Majorana-Weyl representation of the double cover of O(3,3), Spin(3,3), that is being considered. However, following common parlance we neglect this distinction.}. Therefore, the O(3,3) vector field $W^{\alpha}$ is given by a product of Majorana-Weyl spinors $U$ and $V$
$$W^{\alpha} = U^{[\alpha} V^{5]} \propto U^{A}  \gamma^{\alpha}{}_{AB} V^{B},$$
where uppercase Latin letters label spinor indices and take values in $\{1,2,3,5\}.$ Hence, we deduce that 
\begin{equation} 
 \gamma^{\alpha}{}_{AB} \propto \delta^{[\alpha}_{[A} \delta^{5]}_{B]}.
\label{galpha}
\end{equation}
Similarly, if we consider $ W^{ab}= U^{[a} V^{b]}$ for $a= \alpha$ and $b=\beta,$ we deduce that 
\begin{equation}  \tilde{\gamma}_{\alpha}{}_{AB} \propto 
\begin{cases}
\eta_{\alpha A B} &\qquad \textup{for } A,B= 1,2,3 \\
0 & \qquad \textup{otherwise}
\end{cases},
\label{tgalpha}
\end{equation}
where $\eta$ is the alternating symbol. 

We have defined gamma matrices, but we have not shown that they satisfy the Clifford algebra. The $4 \times 4$ O(3,3) $\gamma$-matrices act on Majorana-Weyl spinors, so the corresponding $8 \times 8$ $\Gamma$-matrices, which act on Dirac spinors, are 
\begin{equation} \Gamma^{\Pi}= 
\begin{pmatrix}
0 & (\gamma^{\Pi})^{AB} \\
(\gamma^{\Pi})_{AB} & 0
\end{pmatrix},
\label{Gammadef}
\end{equation}
which satisfy the Clifford algebra
$$\{ \Gamma^{\Pi}, \Gamma^{\Sigma} \}= 2 \eta^{\Pi \Sigma} \mathbb{I}_{8},$$
where
$$ \eta = 
\begin{pmatrix}
0 & \mathbb{I}_{3} \\
\mathbb{I}_{3} & 0
\end{pmatrix}
$$ is the O(3,3) invariant.
In terms of $\gamma$-matrices the Clifford algebra reads 
\begin{equation} 
(\gamma^{\Pi})^{AC} (\gamma^{\Sigma})_{CB} +(\gamma^{\Sigma})^{AC} (\gamma^{\Pi})_{CB} = 2 \eta^{\Pi \Sigma} \delta^{A}_{B}.
\label{Cliffg}
\end{equation}

Using the matrices $\Gamma^{\Pi}= (\Gamma^{\alpha}, \tilde{\Gamma}_{\alpha} ),$ define matrices 
\begin{align}
 \Gamma^{(-) \alpha} &= \frac{1}{\sqrt{2}} ( \Gamma^{\alpha} - \tilde{\Gamma}_{\alpha} ) \notag \\
 \Gamma^{(+) \alpha} &= \frac{1}{\sqrt{2}} ( \Gamma^{\alpha} + \tilde{\Gamma}_{\alpha} ),
\label{gammapm}
\end{align}
which satisfy the Clifford algebra with the diagonal metric, diag(-1,-1,-1,1,1,1), 
\begin{gather*}
 \{ \Gamma^{(-) \alpha},  \Gamma^{(-) \beta} \} = - 2 \delta^{\alpha \beta} \mathbb{I}_{8}, \quad  \{ \Gamma^{(+) \alpha}, \quad \Gamma^{(+) \beta} \} =  2 \delta^{\alpha \beta} \mathbb{I}_{8}, \quad \{ \Gamma^{(-) \alpha},  \Gamma^{(+) \beta} \} = 0.
\end{gather*}
Since the $\Gamma^{(-)}$ matrices square to -1 and $\Gamma^{(+)}$ matrices square to 1, we can assume that the former are antisymmetric while the latter are symmetric. Indeed, choosing the charge conjugation matrix 
$$C= \Gamma^{(-)1}\Gamma^{(-)2}\Gamma^{(-)3},$$ then 
$$ (\Gamma^{(\pm) \alpha})^{T} = - C \Gamma^{(\pm) \alpha} C^{-1}.$$  

Now, from equation \eqref{gammapm} we can deduce that 
$$ (\Gamma^{\alpha})^{T} = \tilde{\Gamma}_{\alpha}, \qquad (\tilde{\Gamma}_{\alpha})^{T} = \Gamma^{\alpha}.$$ 
Hence, looking at equation \eqref{Gammadef}, 
 \begin{equation} 
\begin{pmatrix}
0 & (\gamma^{\alpha})_{BA} \\
(\gamma^{\alpha})^{BA} & 0
\end{pmatrix}
= \begin{pmatrix}
0 & (\tilde{\gamma}_{\alpha})^{AB} \\
(\tilde{\gamma}_{\alpha})_{AB} & 0
\end{pmatrix}. 
\label{grelations}
\end{equation}
The Clifford algebra for the $\gamma$-matrices, equation \eqref{Cliffg}, can now, using the above relation be written as 
 \begin{gather} 
(\gamma^{\alpha})_{CA} (\gamma^{\beta})_{CB} +(\tilde{\gamma}_{\beta})_{CA} (\tilde{\gamma}_{\alpha})_{CB} = 2 \delta_{\alpha}^{\beta} \delta_{AB}, \notag \\
(\tilde{\gamma}_{\alpha})_{CA} (\gamma^{\beta})_{CB} +(\tilde{\gamma}_{\beta})_{CA} (\gamma^{\alpha})_{CB} = 0.
\label{Cliffg2}
\end{gather}
The following gamma-matrices satisfy the Clifford algebra relations above:
\begin{equation} 
 \gamma^{\alpha}{}_{AB} = 2 \sqrt{2} \delta^{[\alpha}_{[A} \delta^{5]}_{B]} \quad \textup{and} \quad  \tilde{\gamma}_{\alpha}{}_{AB} = 
\begin{cases}
\sqrt{2} \eta_{\alpha A B} &\qquad \textup{for } A,B= 1,2,3 \\
0 & \qquad \textup{otherwise}
\end{cases},
\label{Cliffrep}
\end{equation}
which are consistent with equations \eqref{galpha} and \eqref{tgalpha}.

We have shown that under dimensional reduction an SL(5) generalised field in the fundamental representation becomes an O(3,3) spinor field. Therefore, the generalised Lie derivative of a field in the fundamental representation of SL(5), equation \eqref{genLiecovf}, should reduce to the spinorial Lie derivative in O(3,3) \cite{CSW} 
\begin{equation}
\hat{\mathcal{L}}_{X} V = X^{\Pi} \partial_{\Pi} V + \frac{1}{4} (\partial_{\Pi} X_{\Sigma}-\partial_{\Sigma} X_{\Pi}) \Gamma^{\Pi \Sigma} V 
\label{genspinLieO33}.
\end{equation}
To show this we require the expression for $\Gamma^{\Pi \Sigma}$ in the representation in which the dimensionally reduced SL(5) field is related to the O(3,3) spinor, i.e. the representation given in equation \eqref{Cliffrep}. Using equation \eqref{Gammadef}, 
$$\left(\Gamma^{\Pi \Sigma}\right)^{A}{}_{C}= \frac{1}{2} \left( \gamma^{\Pi}{}^{AB} \gamma^{\Sigma}{}_{BC} - \gamma^{\Sigma}{}^{AB} \gamma^{\Pi}{}_{BC} \right).$$ 
The relations in equation \eqref{grelations} can now be used to find $\Gamma^{\Pi \Sigma}$ in the representation of the Clifford algebra given in equation \eqref{Cliffrep}
\begin{gather}
 (\Gamma^{\alpha \beta} )^{A}{}_{B} = 2 \eta^{\alpha \beta A} \delta^{5}_{B}, \qquad (\Gamma_{\alpha \beta} )^{A}{}_{B} = - 2 \eta_{\alpha \beta B} \delta^{A}_{5},\notag \\ (\Gamma^{\alpha}{}_{\beta} )^{A}{}_{B} = \delta^{\alpha}_{\beta} (\delta^{A}_{\gamma} \delta^{\gamma}_{B} - \delta^{A}_{5} \delta^{5}_{B}) - 2  \delta^{\alpha}_{B} \delta^{A}_{\beta}, 
\label{antiG}
\end{gather}
where 
\begin{equation*}
 \Gamma^{\alpha \beta} = \frac{1}{2} (\Gamma^{\alpha} \Gamma^{\beta} - \Gamma^{\beta} \Gamma^{\alpha}), \quad  \Gamma^{\alpha}{}_{\beta} = \frac{1}{2} (\Gamma^{\alpha} \tilde{\Gamma}_{\beta} - \tilde{\Gamma}_{\beta} \Gamma^{\alpha}), \quad \Gamma_{ \alpha \beta} = \frac{1}{2} (\tilde{\Gamma}_{\alpha} \tilde{\Gamma}_{\beta} - \tilde{\Gamma}_{\beta} \tilde{\Gamma}_{\alpha}).
\end{equation*}
Now, we are ready to check the consistency of the two generalised Lie derivatives \eqref{genLiecovf} and \eqref{genspinLieO33}. Inserting equations \eqref{antiG} into the O(3,3) spinorial Lie derivative, \eqref{genspinLieO33}, we obtain 
\begin{align}
(\hat{\mathcal{L}}_{X} V)^A = & X^{\Pi} \partial_{\Pi} V^A + \frac{1}{2} \delta^{A}_{\gamma} (\partial_{\beta} X^{\beta} - \tilde{\partial}^{\beta} \tilde{X}_{\beta} )  V^{\gamma} - \delta^{A}_{\beta} (\partial_{\gamma} X^{\beta} - \tilde{\partial}^{\beta} \tilde{X}_{\gamma} ) V^{\gamma} \notag \\
& \quad + \eta^{A \beta \gamma} (\partial_{\beta} \tilde{X}_{\gamma}) V^{5} - \delta^{A}_{5} \eta_{\beta \gamma \sigma}    (\tilde{\partial}^{\beta} X^{\gamma}) V^{\sigma} -\frac{1}{2} \delta^{A}_{5} ( \partial_{\beta} X^{\beta} -  \tilde{\partial}^{\beta} \tilde{X}_{\beta} ) V^{5}.
\label{spinLie2}
\end{align}
This should be compared to the dimensional reduction of the generalised Lie derivative on a field in the fundamental of SL(5). We use the same reduction ansatz as before, namely that derivatives along the fourth direction vanish, and evaluate the Lie derivative, equation \eqref{genLiecovf}, for the index $a= \alpha$ and $a=5$, respectively,
\begin{align*}
 (\hat{\mathcal{L}}_{X} V)^{\alpha} = & X^{\Pi} \partial_{\Pi} V^{\alpha} + \frac{1}{2} (\partial_{\beta} X^{\beta} - \tilde{\partial}^{\beta} \tilde{X}_{\beta} ) V^{\alpha} \\
& \qquad \qquad \qquad - (\partial_{\beta} X^{\alpha} - \tilde{\partial}^{\alpha} \tilde{X}_{\beta} ) V^{\beta} + \eta^{\alpha\beta \gamma} (\partial_{\beta} \tilde{X}_{\gamma}) V^{5},\\
(\hat{\mathcal{L}}_{X} V)5 = & X^{\Pi} \partial_{\Pi} V5 - \eta_{\beta \gamma \sigma} (\tilde{\partial}^{\beta} X^{\gamma}) V^{\sigma} -\frac{1}{2} ( \partial_{\beta} X^{\beta} -  \tilde{\partial}^{\beta} \tilde{X}_{\beta} ) V^{5}.
\end{align*}
Comparing the above equations with \eqref{spinLie2}, we conclude that the dimensionally reduced SL(5) generalised Lie derivative acting on a field in the fundamental representation is equal to the O(3,3) generalised spinorial Lie derivative.

In this section, we have shown that under dimensional reduction the SL(5) generalised field in the antisymmetric representation becomes an O(3,3) vector, while an SL(5) vector field becomes an O(3,3) spinor. Furthermore, by dimensionally reducing the SL(5) generalised Lie derivatives, \eqref{genLiecov} and \eqref{genLiecovf}, we find the O(3,3) generalised Lie derivatives on vector fields and spinors. The equality between these objects was established without the use of any section condition. However, the section condition seems to play an integral role in generalised geometry. In particular, applications of generalised geometry to physics rely on specific solutions of the section condition. For example, the rewriting of supergravity actions in terms of a generalised metric uses a particular solution in which the fields only depend on the spacetime coordinates and not on the brane windings. In the next section, we will consider the section condition, giving examples of other solutions where the fields depend on brane windings.

\section{Section condition}
\label{sectioncon}

The section condition for the SL(5) generalised geometry,
\begin{equation}
  \partial_{[ab} \partial_{cd]} X =0 \qquad \qquad  \textup{on all fields }X,
\label{section}
\end{equation}
 was found in section \ref{geo} by requiring closure of the algebra of generalised diffeomorphisms. The second equation in \eqref{seccon} is necessarily satisfied given the first. The section condition can be dimensionally reduced to obtain the O$(3,3)$ section condition
$$ \tilde{\partial}^{\alpha} \partial_{\alpha} =0.$$
Consider the operator constraint
$$\eta^{abcde} \partial_{bc} \partial_{de} =0.$$ 
Reducing this along the fourth direction with the ansatz that derivative along this direction vanishes, the section condition becomes
$$\eta^{4 \alpha \beta \gamma 5} \partial_{\alpha \beta} \partial_{\gamma} =0, $$
which implies the O(3,3) section condition 
$$ \tilde{\partial}^{\alpha} \partial_{\alpha}=0.$$
The O$(d,d)$ section condition is a Laplace equation in a Kleinian space. The general solution to this type of differential equation was found in \cite{Osoln} in the context of $\mathcal{N}=2$ strings.    

One can use the Weyl group to investigate solutions to equation \eqref{section}. The Weyl group permutes the coordinates into each other. For the SL(5) group the Weyl group is the permutation group of 5 elements $S_{5}.$ Consider the Fourier transform of the section condition \eqref{section}, which implies that
\begin{equation}
 p_{[ab} p_{cd]}=0.
\label{fsection}
\end{equation}
Since the Weyl group takes into account the redundancy in our labelling of coordinates, pick the indices in the above equation to be $ 1, \dots, 4.$ Therefore, equation \eqref{fsection} becomes
\begin{equation}
\mathbf{p}^{T} 
\begin{pmatrix}
 0 & 0 & 0 & 0 & 0 & 1 \\
 0 & 0 & 0 & 0 & -1 & 0 \\
 0 & 0 & 0 & 1 & 0 & 0 \\
 0 & 0 & 1 & 0 & 0 & 0 \\
 0 & -1 & 0 & 0 & 0 & 0 \\
 1 & 0 & 0 & 0 & 0 & 0 
\end{pmatrix} \mathbf{p}=0,
\label{peqn}
\end{equation}
where $\mathbf{p}=(p_{12},p_{13},p_{14},p_{23},p_{24},p_{34}).$ The matrix above can be diagonalised so that the above equation reads 
\begin{equation}
 a^2 + b^2 + c^2 - j^2 - k^2 - l^2=0,
\label{deqn}
\end{equation}
where 
\begin{gather}
       a= p_{12} + p_{34} , \qquad b= p_{13} + p_{24} , \qquad c= p_{14} + p_{23} \notag \\
j= p_{12} - p_{34} , \qquad k= p_{13} - p_{24} , \qquad l= p_{14} - p_{23}.
\end{gather}
Hence the SL(5) section condition is also the Laplace equation on a Kleinian space. Consider now the other 4 equations in \eqref{fsection}. These equations can be written as a matrix equation
$$N \mathbf{p}^{(5)} =0,$$
where $\mathbf{p}^{(5)} = (p_{15}, p_{25}, p_{35}, p_{45}),$ and 
$$N=
\begin{pmatrix}
0 & p_{34} & - p_{24} & p_{23} \\
- p_{34} & 0 & p_{14} & - p_{13} \\
p_{24} & - p_{14} & 0 & p_{12} \\
- p_{23} &  p_{13} & - p_{12} & 0 \\
\end{pmatrix}.
$$
If we think of $\mathbf{p}$ as solving equation \eqref{peqn}, then the determinant of the matrix $N$ vanishes, so it is necessarily degenerate. We can use the Gauss elimination method to find the rank of the matrix. Assuming that it has non-zero rank, we can without loss of generality take $p_{34} \neq 0,$ in which case the matrix $N$ reduces to its row echelon form
\begin{equation}
 \begin{pmatrix}
1 & 0 & - p_{14}/p_{34} & p_{13}/p_{34} \\
0 & 1 & - p_{24}/p_{34} & p_{23}/p_{34} \\
0 & 0 & p_{34} p_{12} - p_{24} p_{13} + p_{23} p_{24} & 0 \\
0 & 0 & 0 & p_{34} p_{12} - p_{24} p_{13} + p_{23} p_{24} \\
\end{pmatrix}.
\end{equation}
The expression $p_{34} p_{12} - p_{24} p_{13} + p_{23} p_{24}$ vanishes by equation \eqref{peqn}, hence the rank of  $N$ is less than or equal to two. But because the matrix $N$ is antisymmetric it cannot be of rank one. Therefore, $N$ has rank zero or two. The former case corresponds to letting the fields be independent of the winding coordinates. This is the section condition that was used in \cite{davidmalcolm} to recover the supergravity action from the duality invariant formulation in terms of the generalised metric. The latter case, when the rank of $N$ is two, gives alternative section conditions where the fields can depend on winding coordinates. For example, the choice $$p_{12}, p_{23}, p_{13}, p_{15}, p_{25}, p_{35}=0$$ solves the SL(5) condition, and so the fields depend on the coordinates $$ x^{4}, y_{14}, y_{24}, y_{34}.$$

It is possible to consider alternative solutions of the section condition and find the theory to which the duality-invariant formulation reduces. This leads to different duality frames of eleven-dimensional supergravity.

\section{Discussion}

In this paper, we constructed the generalised geometry associated to the SL(5) duality group. This is the duality group that appears on the reduction of eleven-dimensional supergravity to seven dimensions. An SL(5) covariant generalised Lie derivative was constructed by dualising the winding coordinates. The generalised Lie derivative on a field in the antisymmetric and vector representations was shown to give the generalised Lie derivative on an O(3,3) vector and spinor. Therefore, an SL(5) vector field becomes an O(3,3) spinor under dimensional reduction. This is perhaps related to the fact that 3-form potential in eleven-dimensional supergravity gives rise to both NSNS and RR fields under dimensional reduction. 

The generalised Lie derivative generates generalised diffeomorphisms which encode the diffeomorphisms and U(1) gauge transformations. The closure of the algebra of generalised diffeomorphisms is only satisfied up to a constraint, which we identify with the section condition. The section condition has solutions whereby the fields can also depend on the winding coordinates. However, unlike the O$(d,d)$ section condition the SL(5) section condition does not admit solutions whereby the fields only depend on winding coordinates. An interesting prospect to explore is to reduce the duality-invariant dynamics with respect to these alternative section choices. These different section choices will lead to different duality frames for the theory and allow for the definition of non-geometries based on transition functions that are given by SL(5) transformations.

Furthermore, it will be interesting to extend this work to the other dualities of M-theory. It is likely that the section conditions for the larger duality groups, $SO(5,5), E_{6}, E_{7}$ and $E_{8},$ can also be found by imposing closure of the algebra of generalised diffeomorphisms. 

The section condition is a quantum off-shell condition rather than an on-shell condition. Equation \eqref{fsection} must be read as a quantum mechanical condition as $p= i \hbar \frac{\partial}{\partial x}.$ 

The type of constraint given by the section condition has been observed before in the context of half-BPS states and U-duality {\cite{OP},\cite{west04a}}. It would be of great interest to see if this connection remains true for the larger duality groups.

\paragraph*{Acknowledgements}
We would like to thank Mihalis Dafermos, Gary Gibbons, Chris Hull, Boris Pioline and Peter West for discussions. DSB is supported in part by the Queen Mary STFC rolling grant ST/G000565/1. HG is supported by an STFC grant. MG is supported by an EPSRC grant. HG and MG thank St. John's College Cambridge for their support. MJP is in
part supported by the STFC rolling grant STJ000434/1. MJP would like to thank the Mitchell foundation and Trinity College Cambridge for their generous support.

\appendix

\section{Jacobiator of the generalised Lie bracket} \label{Jac}

In this appendix, we consider the status of the Jacobi identity for the generalised Lie bracket, equation \eqref{genbra}.

Define the Jacobiator of the generalised Lie bracket to be
\begin{equation}
 J(X,Y,Z)^{ab} = \left( [[X,Y]_{G},Z]_G + [[Y,Z]_{G},X]_G + [[Z,X]_{G},Y]_G \right)^{ab},
\end{equation}
where the bracket $[\;,\;]_G$ is defined in equation \eqref{genbra} and $X$, $Y$, $Z$ are SL(5) bivectors, i.e. they have index structure $X^{ab}=X^{[ab]}$.  The Jacobi identity for the generalised Lie bracket would be
\begin{equation}
 J(X,Y,Z)^{ab}=0.
\end{equation}

Using the fact that the generalised Lie bracket is given by antisymmetrising a generalised Lie derivative, equation \eqref{defC}, i.e.
\begin{equation}
[X,Y]_{G}=\frac{1}{2}(\hat{\mathcal{L}}_{X} Y - \hat{\mathcal{L}}_{Y} X),
\end{equation}
we rewrite the Jacobiator
\begin{align}
  J(X,Y,Z)^{ab} =& \frac{1}{4}([\hat{\mathcal{L}}_{X}, \hat{\mathcal{L}}_{Y}] Z + [\hat{\mathcal{L}}_{Y}, \hat{\mathcal{L}}_{Z}] X + [\hat{\mathcal{L}}_{Z}, \hat{\mathcal{L}}_{X}] Y)^{ab} \notag \\
&- \frac{1}{2} (\hat{\mathcal{L}}_{[X,Y]_G} Z + \hat{\mathcal{L}}_{[Y,Z]_G} X + \hat{\mathcal{L}}_{[Z,X]_G} Y)^{ab}.
\end{align}
Using, equation \eqref{antiLie}, the above expression reduces to
\begin{equation}
 J(X,Y,Z)^{ab} =- \frac{1}{4} (\hat{\mathcal{L}}_{[X,Y]_G} Z + \hat{\mathcal{L}}_{[Y,Z]_G} X + \hat{\mathcal{L}}_{[Z,X]_G} Y)^{ab} + \ldots,
\end{equation}
where the ellipses here and below denote terms that vanish if the section condition is assumed to hold.

Expanding out the terms above using the definition of the generalised Lie derivative given in equation \eqref{genLiecov}, we find that
\begin{align}
 J(X,Y,Z)^{ab} =& - \frac{1}{8} \left\lbrace \left( Z^{ab} \partial_{cd} [X,Y]^{cd}_{G} + 4 Z^{c[a} \partial_{cd} [X,Y]^{b]d}_G \right) \right. \notag \\
&\left. + \left( X^{ab} \partial_{cd} [Y,Z]^{cd}_{G} + 4 X^{c[a} \partial_{cd} [Y,Z]^{b]d}_G \right) 
+ \left( Y^{ab} \partial_{cd} [Z,X]^{cd}_{G} + 4 Y^{c[a} \partial_{cd} [Z,X]^{b]d}_G \right) \right. \notag \\
&\left. + [X,Y]^{cd}_{G} \partial_{cd} Z^{ab} + [Y,Z]^{cd}_{G} \partial_{cd} X^{ab} + [Z,X]^{cd}_{G} \partial_{cd} Y^{ab}  \right\rbrace + \ldots.
\label{Jexp}
\end{align}
$J(X,Y,Z)^{ab}$ does not vanish even if the section condition is assumed.
However, one can show that the Jacobi identity is satisfied when acting on fields modulo terms that vanish by the section condition.  Using the following identity
\begin{equation*}
6 Z^{ab} \partial_{[cd} [X,Y]^{cd}_G \partial_{ab]} F = \left\lbrace \left( Z^{ab} \partial_{cd} [X,Y]^{cd}_{G} + 4 Z^{ca} \partial_{cd} [X,Y]^{bd}_G \right) + Z^{cd} \partial_{cd} [X,Y]^{ab}_{G} \right\rbrace \partial_{ab} F
\end{equation*}
repeatedly in the expression for $J(X,Y,Z)^{ab}$ above, \eqref{Jexp}, gives
\begin{align}
 J(X,Y,Z)^{ab} \partial_{ab} F = & \frac{1}{8} \left\lbrace Z^{cd} \partial_{cd} [X,Y]^{ab}_{G} + X^{cd} \partial_{cd} [Y,Z]^{ab}_{G} + Y^{cd} \partial_{cd} [Z,X]^{ab}_{G} \right. \notag \\
&\left. - [X,Y]^{cd}_{G} \partial_{cd} Z^{ab} - [Y,Z]^{cd}_{G} \partial_{cd} X^{ab} - [Z,X]^{cd}_{G} \partial_{cd} Y^{ab}  \right\rbrace \partial_{ab} F + \ldots,
\end{align}
where $F$ is any SL(5) covariant object, where indices have been suppressed.
Finally, using the operator identity 
\begin{equation*}
 [X,Y]^{cd}_{G} \partial_{cd} = \frac{1}{2} (X^{ef} \partial_{ef} Y^{cd} - Y^{ef} \partial_{ef} X^{cd}) \partial_{cd} + \ldots,
\end{equation*}
to simplify all six terms in the expression for $J(X,Y,Z)^{ab} \partial_{ab} F$ above allows us to conclude that
\begin{equation}
 J(X,Y,Z)^{ab} \partial_{ab} F = 0,
\end{equation}
modulo terms that vanish by the section condition.

\bibliography{M}

\providecommand{\href}[2]{#2}\begingroup\raggedright\begin{thebibliography}{10}

\bibitem{SO(8)}
E.~Cremmer and B.~Julia, ``{The SO(8) Supergravity},''
  \href{http://dx.doi.org/10.1016/0550-3213(79)90331-6}{{\em Nucl.Phys.} {\bf
  B159} (1979)  141}.

\bibitem{julia}
B.~Julia, ``Group disintegrations,'' in {\em Superspace and Supergravity:
  Proceedings of the Nuffield Workshop, Cambridge 1980}, S.~W. Hawking and
  M.~Rocek, eds., pp.~331--350.
\newblock Cambridge University Press, 1981.

\bibitem{TM}
J.~Thierry-Mieg and B.~Morel, ``Superalgebras in exceptional gravity,'' in {\em
  Superspace and Supergravity: Proceedings of the Nuffield Workshop, Cambridge
  1980}, S.~W. Hawking and M.~Rocek, eds., pp.~351--363.
\newblock Cambridge University Press, 1981.

\bibitem{nicolaidewit}
B.~de~Wit and H.~Nicolai, ``d = 11 supergravity with local {SU}(8)
  invariance,'' \href{http://dx.doi.org/10.1016/0550-3213(86)90290-7}{{\em
  Nucl.Phys.} {\bf B274} (1986)  363}.

\bibitem{nicolia87}
H.~Nicolai, ``D = 11 supergravity with local {SO}(16) invariance,''
  \href{http://dx.doi.org/10.1016/0370-2693(87)91102-6}{{\em Phys.Lett.} {\bf
  B187} (1987)  316}.

\bibitem{weste11}
P.~C. West, ``{E(11) and M theory},''
  \href{http://dx.doi.org/10.1088/0264-9381/18/21/305}{{\em Class.Quant.Grav.}
  {\bf 18} (2001)  4443--4460}, \href{http://arxiv.org/abs/hep-th/0104081}{{\tt
  arXiv:hep-th/0104081 [hep-th]}}.

\bibitem{dft1}
C.~Hull and B.~Zwiebach, ``{Double Field Theory},''
  \href{http://dx.doi.org/10.1088/1126-6708/2009/09/099}{{\em JHEP} {\bf 0909}
  (2009)  099}, \href{http://arxiv.org/abs/0904.4664}{{\tt arXiv:0904.4664
  [hep-th]}}.

\bibitem{dft2}
C.~Hull and B.~Zwiebach, ``{The Gauge algebra of double field theory and
  Courant brackets},''
  \href{http://dx.doi.org/10.1088/1126-6708/2009/09/090}{{\em JHEP} {\bf 0909}
  (2009)  090}, \href{http://arxiv.org/abs/0908.1792}{{\tt arXiv:0908.1792
  [hep-th]}}.

\bibitem{dft3}
O.~Hohm, C.~Hull, and B.~Zwiebach, ``{Background independent action for double
  field theory},'' \href{http://dx.doi.org/10.1007/JHEP07(2010)016}{{\em JHEP}
  {\bf 1007} (2010)  016}, \href{http://arxiv.org/abs/1003.5027}{{\tt
  arXiv:1003.5027 [hep-th]}}.

\bibitem{genmetstring}
O.~Hohm, C.~Hull, and B.~Zwiebach, ``{Generalized metric formulation of double
  field theory},'' \href{http://dx.doi.org/10.1007/JHEP08(2010)008}{{\em JHEP}
  {\bf 1008} (2010)  008}, \href{http://arxiv.org/abs/1006.4823}{{\tt
  arXiv:1006.4823 [hep-th]}}.

\bibitem{hitchin1}
N.~Hitchin, ``{Generalized Calabi-Yau manifolds},''
  \href{http://dx.doi.org/10.1093/qjmath/54.3.281}{{\em Quart.J.Math.Oxford
  Ser.} {\bf 54} (2003)  281--308},
  \href{http://arxiv.org/abs/math/0209099}{{\tt arXiv:math/0209099 [math-dg]}}.

\bibitem{hitchin2}
N.~Hitchin, ``{Brackets, forms and invariant functionals},''
  \href{http://arxiv.org/abs/math/0508618}{{\tt arXiv:math/0508618 [math-dg]}}.
  dedicated to the memory of Shiing-Shen Chern.

\bibitem{gualtieri}
M.~Gualtieri, ``{Generalized complex geometry},''
  \href{http://arxiv.org/abs/math/0401221}{{\tt arXiv:math/0401221 [math-dg]}}.
  Ph.D. Thesis (Advisor: Nigel Hitchin).

\bibitem{courant}
T.~Courant, ``{Dirac manifolds},'' {\em Trans. Amer. Math. Soc.} {\bf 319}
  (1990)  631--661.

\bibitem{KZ}
T.~Kugo and B.~Zwiebach, ``{Target space duality as a symmetry of string field
  theory},'' \href{http://dx.doi.org/10.1143/PTP.87.801}{{\em Prog.Theor.Phys.}
  {\bf 87} (1992)  801--860}, \href{http://arxiv.org/abs/hep-th/9201040}{{\tt
  arXiv:hep-th/9201040 [hep-th]}}.

\bibitem{siegel1}
W.~Siegel, ``{Two vierbein formalism for string inspired axionic gravity},''
  \href{http://dx.doi.org/10.1103/PhysRevD.47.5453}{{\em Phys.Rev.} {\bf D47}
  (1993)  5453--5459}, \href{http://arxiv.org/abs/hep-th/9302036}{{\tt
  arXiv:hep-th/9302036 [hep-th]}}.

\bibitem{siegel2}
W.~Siegel, ``{Superspace duality in low-energy superstrings},''
  \href{http://dx.doi.org/10.1103/PhysRevD.48.2826}{{\em Phys.Rev.} {\bf D48}
  (1993)  2826--2837}, \href{http://arxiv.org/abs/hep-th/9305073}{{\tt
  arXiv:hep-th/9305073 [hep-th]}}.

\bibitem{hullm}
C.~Hull, ``{Generalised Geometry for M-Theory},''
  \href{http://dx.doi.org/10.1088/1126-6708/2007/07/079}{{\em JHEP} {\bf 0707}
  (2007)  079}, \href{http://arxiv.org/abs/hep-th/0701203}{{\tt
  arXiv:hep-th/0701203 [hep-th]}}.

\bibitem{PW}
P.~P. Pacheco and D.~Waldram, ``{M-theory, exceptional generalised geometry and
  superpotentials},''
  \href{http://dx.doi.org/10.1088/1126-6708/2008/09/123}{{\em JHEP} {\bf 0809}
  (2008)  123}, \href{http://arxiv.org/abs/0804.1362}{{\tt arXiv:0804.1362
  [hep-th]}}.

\bibitem{hillmann}
C.~Hillmann, ``{Generalized E(7(7)) coset dynamics and D=11 supergravity},''
  \href{http://dx.doi.org/10.1088/1126-6708/2009/03/135}{{\em JHEP} {\bf 0903}
  (2009)  135}, \href{http://arxiv.org/abs/0901.1581}{{\tt arXiv:0901.1581
  [hep-th]}}.

\bibitem{davidmalcolm}
D.~S. Berman and M.~J. Perry, ``{Generalized Geometry and M theory},''
  \href{http://dx.doi.org/10.1007/JHEP06(2011)074}{{\em JHEP} {\bf 1106} (2011)
   074}, \href{http://arxiv.org/abs/1008.1763}{{\tt arXiv:1008.1763 [hep-th]}}.

\bibitem{BGP}
D.~S. Berman, H.~Godazgar, and M.~J. Perry, ``{SO(5,5) duality in M-theory and
  generalized geometry},''
  \href{http://dx.doi.org/10.1016/j.physletb.2011.04.046}{{\em Phys. Lett.}
  {\bf B700} (2011)  65--67},
\href{http://arxiv.org/abs/1103.5733}{{\tt arXiv:1103.5733 [hep-th]}}.

\bibitem{BGPW1}
D.~S. Berman, H.~Godazgar, M.~J. Perry, and P.~C. West, ``{Exceptional geometry
  for M-theory},''. In preparation.

\bibitem{BGPW2}
D.~S. Berman, H.~Godazgar, M.~J. Perry, and P.~C. West, ``{Exceptional geometry
  from $E_{11}$},''. In preparation.

\bibitem{westl1}
P.~C. West, ``{E(11), SL(32) and central charges},''
  \href{http://dx.doi.org/10.1016/j.physletb.2003.09.059}{{\em Phys.Lett.} {\bf
  B575} (2003)  333--342}, \href{http://arxiv.org/abs/hep-th/0307098}{{\tt
  arXiv:hep-th/0307098 [hep-th]}}.

\bibitem{KW03}
A.~Kleinschmidt and P.~C. West, ``{Representations of G+++ and the role of
  space-time},'' \href{http://dx.doi.org/10.1088/1126-6708/2004/02/033}{{\em
  JHEP} {\bf 0402} (2004)  033},
  \href{http://arxiv.org/abs/hep-th/0312247}{{\tt arXiv:hep-th/0312247
  [hep-th]}}.

\bibitem{west04a}
P.~C. West, ``{E(11) origin of brane charges and U-duality multiplets},''
  \href{http://dx.doi.org/10.1088/1126-6708/2004/08/052}{{\em JHEP} {\bf 0408}
  (2004)  052}, \href{http://arxiv.org/abs/hep-th/0406150}{{\tt
  arXiv:hep-th/0406150 [hep-th]}}.

\bibitem{west04b}
P.~C. West, ``{Brane dynamics, central charges and E(11)},''
  \href{http://dx.doi.org/10.1088/1126-6708/2005/03/077}{{\em JHEP} {\bf 0503}
  (2005)  077}, \href{http://arxiv.org/abs/hep-th/0412336}{{\tt
  arXiv:hep-th/0412336 [hep-th]}}.

\bibitem{west10}
P.~West, ``{Generalised space-time and duality},''
  \href{http://dx.doi.org/10.1016/j.physletb.2010.08.054}{{\em Phys.Lett.} {\bf
  B693} (2010)  373--379}, \href{http://arxiv.org/abs/1006.0893}{{\tt
  arXiv:1006.0893 [hep-th]}}.

\bibitem{westiia}
P.~West, ``{$E_{11}$, generalised space-time and IIA string theory},''
  \href{http://dx.doi.org/10.1016/j.physletb.2010.12.041}{{\em Phys.Lett.} {\bf
  B696} (2011)  403--409}, \href{http://arxiv.org/abs/1009.2624}{{\tt
  arXiv:1009.2624 [hep-th]}}.

\bibitem{relSM}
D.~C. Thompson, ``{Duality Invariance: From M-theory to Double Field Theory},''
  \href{http://dx.doi.org/10.1007/JHEP08(2011)125}{{\em JHEP} {\bf 1108} (2011)
   125}, \href{http://arxiv.org/abs/1106.4036}{{\tt arXiv:1106.4036 [hep-th]}}.

\bibitem{dhet}
O.~Hohm and S.~K. Kwak, ``{Double Field Theory Formulation of Heterotic
  Strings},'' \href{http://dx.doi.org/10.1007/JHEP06(2011)096}{{\em JHEP} {\bf
  1106} (2011)  096}, \href{http://arxiv.org/abs/1103.2136}{{\tt
  arXiv:1103.2136 [hep-th]}}.

\bibitem{dtypeii}
O.~Hohm, S.~K. Kwak, and B.~Zwiebach, ``{Unification of Type II Strings and
  T-duality},'' \href{http://arxiv.org/abs/1106.5452}{{\tt arXiv:1106.5452
  [hep-th]}}.

\bibitem{dtypeiim}
O.~Hohm and S.~K. Kwak, ``{Massive Type II in Double Field Theory},''
  \href{http://arxiv.org/abs/1108.4937}{{\tt arXiv:1108.4937 [hep-th]}}.

\bibitem{CSW}
A.~Coimbra, C.~Strickland-Constable, and D.~Waldram, ``{Supergravity as
  Generalised Geometry I: Type II Theories},''
  \href{http://arxiv.org/abs/1107.1733}{{\tt arXiv:1107.1733 [hep-th]}}.

\bibitem{Albertsson:2008gq}
C.~Albertsson, T.~Kimura, and R.~A. Reid-Edwards, ``{D-branes and doubled
  geometry},'' \href{http://dx.doi.org/10.1088/1126-6708/2009/04/113}{{\em
  JHEP} {\bf 0904} (2009)  113}, \href{http://arxiv.org/abs/0806.1783}{{\tt
  arXiv:0806.1783 [hep-th]}}.

\bibitem{Jeon:2010rw}
I.~Jeon, K.~Lee, and J.-H. Park, ``{Differential geometry with a projection:
  Application to double field theory},''
  \href{http://dx.doi.org/10.1007/JHEP04(2011)014}{{\em JHEP} {\bf 1104} (2011)
   014}, \href{http://arxiv.org/abs/1011.1324}{{\tt arXiv:1011.1324 [hep-th]}}.

\bibitem{Hohm:2010xe}
O.~Hohm and S.~K. Kwak, ``{Frame-like Geometry of Double Field Theory},''
  \href{http://dx.doi.org/10.1088/1751-8113/44/8/085404}{{\em J.Phys.A} {\bf
  A44} (2011)  085404}, \href{http://arxiv.org/abs/1011.4101}{{\tt
  arXiv:1011.4101 [hep-th]}}.

\bibitem{Jeon:2011kp}
I.~Jeon, K.~Lee, and J.-H. Park, ``{Double field formulation of Yang-Mills
  theory},'' \href{http://dx.doi.org/10.1016/j.physletb.2011.05.051}{{\em
  Phys.Lett.} {\bf B701} (2011)  260--264},
  \href{http://arxiv.org/abs/1102.0419}{{\tt arXiv:1102.0419 [hep-th]}}.

\bibitem{Jeon:2011cn}
I.~Jeon, K.~Lee, and J.-H. Park, ``{Stringy differential geometry, beyond
  Riemann},'' \href{http://dx.doi.org/10.1103/PhysRevD.84.044022}{{\em
  Phys.Rev.} {\bf D84} (2011)  044022},
  \href{http://arxiv.org/abs/1105.6294}{{\tt arXiv:1105.6294 [hep-th]}}.

\bibitem{Andriot:2011uh}
D.~Andriot, M.~Larfors, D.~Lust, and P.~Patalong, ``{A ten-dimensional action
  for non-geometric fluxes},''
  \href{http://dx.doi.org/10.1007/JHEP09(2011)134}{{\em JHEP} {\bf 1109} (2011)
   134}, \href{http://arxiv.org/abs/1106.4015}{{\tt arXiv:1106.4015 [hep-th]}}.

\bibitem{Albertsson:2011ux}
C.~Albertsson, S.-H. Dai, P.-W. Kao, and F.-L. Lin, ``{Double Field Theory for
  Double D-branes},'' \href{http://dx.doi.org/10.1007/JHEP09(2011)025}{{\em
  JHEP} {\bf 1109} (2011)  025}, \href{http://arxiv.org/abs/1107.0876}{{\tt
  arXiv:1107.0876 [hep-th]}}.

\bibitem{Kan:2011vg}
N.~Kan, K.~Kobayashi, and K.~Shiraishi, ``{Equations of Motion in Double Field
  Theory: From particles to scale factors},''
  \href{http://arxiv.org/abs/1108.5795}{{\tt arXiv:1108.5795 [hep-th]}}.

\bibitem{Aldazabal:2011nj}
G.~Aldazabal, W.~Baron, D.~Marques, and C.~Nunez, ``{The effective action of
  Double Field Theory},'' \href{http://arxiv.org/abs/1109.0290}{{\tt
  arXiv:1109.0290 [hep-th]}}.

\bibitem{Jeon:2011vx}
I.~Jeon, K.~Lee, and J.-H. Park, ``{Incorporation of fermions into double field
  theory},'' \href{http://arxiv.org/abs/1109.2035}{{\tt arXiv:1109.2035
  [hep-th]}}.

\bibitem{boundary}
D.~S. Berman, E.~T. Musaev, and M.~J. Perry, ``{Boundary Terms in Generalized
  Geometry and doubled field theory},''
  \href{http://arxiv.org/abs/1110.3097}{{\tt arXiv:1110.3097 [hep-th]}}.

\bibitem{GMPW}
M.~Grana, R.~Minasian, M.~Petrini, and D.~Waldram, ``{T-duality, Generalized
  Geometry and Non-Geometric Backgrounds},''
  \href{http://dx.doi.org/10.1088/1126-6708/2009/04/075}{{\em JHEP} {\bf 0904}
  (2009)  075}, \href{http://arxiv.org/abs/0807.4527}{{\tt arXiv:0807.4527
  [hep-th]}}.

\bibitem{Osoln}
J.~W. Barrett, G.~Gibbons, M.~Perry, C.~Pope, and P.~Ruback, ``{Kleinian
  geometry and the N=2 superstring},''
  \href{http://dx.doi.org/10.1142/S0217751X94000650}{{\em Int.J.Mod.Phys.} {\bf
  A9} (1994)  1457--1494}, \href{http://arxiv.org/abs/hep-th/9302073}{{\tt
  arXiv:hep-th/9302073 [hep-th]}}.

\bibitem{OP}
N.~Obers and B.~Pioline, ``{U duality and M theory},''
  \href{http://dx.doi.org/10.1016/S0370-1573(99)00004-6}{{\em Phys.Rept.} {\bf
  318} (1999)  113--225}, \href{http://arxiv.org/abs/hep-th/9809039}{{\tt
  arXiv:hep-th/9809039 [hep-th]}}.

\end{thebibliography}\endgroup
\bibliographystyle{utphys}
\end{document}